\documentclass[preprint]{elsarticle}
\usepackage{amssymb}
\usepackage{epsfig}
\usepackage{lineno}

\journal{Nuclear Instruments and Methods in Physics Research A}

\begin{document}

\begin{frontmatter}

\title{Determination of shower central position in laterally segmented lead-fluoride electromagnetic calorimeters}

\author[label1]{M.~Mazouz\corref{cor1}}
\author[label1]{L.~Ghedira}
\author[label2]{E.~Voutier}

\address[label1]{Facult\'e des Sciences de Monastir\\ D\'epartement de Physique\\ Monastir 5000, Tunisia \\ \phantom{EmptyLine}}

\address[label2]{Institut de Physique Nucl\'eaire d'Orsay\\ IN2P3/CNRS, Universit\'e Paris Sud \\ 
15 rue Georges Cl\'emenceau, 91406 Orsay, France}

\cortext[cor1]{corresponding author : mazouz@jlab.org}


\begin{abstract}

The spatial resolution of laterally segmented electromagnetic calorimeters is studied on the basis of Monte-Carlo simulations 
worked-out for lead fluoride material. Parametrization of the relative resolution is proposed and optimized in terms of the 
energy of incoming particles and the elementary size of the calorimeter blocks. A new fit algorithm method is proposed that 
improves spatial resolution at high energies, and provides guidance for the design optimization of electromagnetic calorimeters.  
\end{abstract}

\begin{keyword}
calorimeter \sep electromagnetic shower \sep impact position \sep spatial resolution \sep lead fluoride
\end{keyword}


\end{frontmatter}


\hyphenation{pa-ra-me-tri-za-tion}

\section{Introduction}
\label{intro}

Electromagnetic calorimeters are fundamental elements of numerous experiments ranging from nuclear to hadronic and high-energy 
physics. Their main goal is to measure precisely the energy of the detected electrons and photons. The depth of the calorimeters 
is generally taken to be large enough to avoid energy leaks and to allow a full development of the electromagnetic shower created 
by the incident  particles. In addition to the energy measurement, the impact position of particles, corresponding to the shower 
central position in case of normal incidence, is usually required in order to provide refined information as particle identification. 
Indeed, the knowledge of the photon or the electron energy and both the impact and interaction vertex positions allow the determination 
of the particle four-vector. This is of direct relevance for instance in the experimental determination of the origin of two photon 
events from the decay of $\pi^0$- or $\eta$-mesons where the 2$\gamma$-invariant mass allows identifying the meson nature. It has been 
shown is some experiments, that the spatial resolution is of equal importance than the energy resolution for particle  identification~\cite{ref1}.
 
The knowledge of the particle impact position requires a laterally segmented calorimeter where the energy is released in a cluster 
of adjacent blocks. The impact position can then be determined using the energy deposited in each block. Laterally  segmented 
calorimeters such as lead-fluoride (PbF$_2$) are frequently employed to determine the energy and the position of the showering particle~\cite{{ref2},{ref3},{ref4}}. This high-density material has a short radiation length ($X_{0}=0.93$ cm) and a small Moli\`ere 
radius ($r_{M}=2.12$ cm) leading to compact detector geometries~\cite{{ref5},{ref14}}. 

Many efforts are still developed for optimizing the performances of such lead-fluoride calorimeters, in particular their energy and 
spatial resolution~\cite{ref16}. We investigate in this work, on the basis of GEANT4 Monte Carlo simulations~\cite{ref6}, the spatial resolution 
of a PbF$_2$ calorimeter for several block sizes and different particle energies. Two approaches are successively considered, discussed, and optimized to obtain a parametrization of the spatial resolution in terms of the block size and the particle energy. 
Finally, these approaches are compared and the effect of the energy resolution on the spatial resolution is addressed. 


\section{Simulation}
\label{sec2}

\begin{figure}[h!]
    \begin{center}
      \epsfig{file=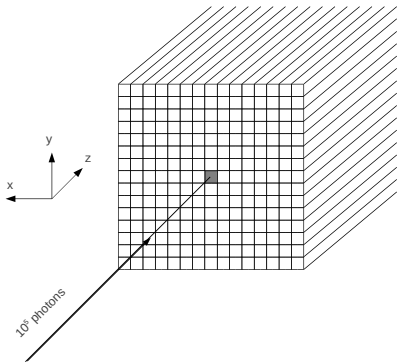, width=5.5cm, angle =-90}
\end{center}
\caption{Schematic of the geometry of the simulated calorimeter.}
\label{fig1}
\end{figure}
The electromagnetic calorimeter is simulated within the GEANT4 framework following a $15 \times 15$ PbF$_2$ block matrix.  
Each block has a transverse square shape with dimensions $d \times d$~cm$^2$ and large enough depth (50$X_{0}$) along 
the $z$-axis to avoid energy leaks. Different configurations corresponding to 13 different transverse size varying from 
$0.3 r_{M}$ to $5 r_{M}$ are studied. The minimum block size ($0.3 r_{M}$) is consistent with a large enough minimal calorimeter 
($4.5 r_M$$\times$$4.5 r_M$) to contain the electromagnetic shower. For each configuration, photons of energy $E$ are sent parallel to 
the $z$-axis and impinge normally on the central region of the calorimeter (Fig.~\ref{fig1}). The known impact position ($x$,$y$) 
of incident particles is distributed uniformly over the area of the central block $(-d/2,-d/2) < (x,y) < (d/2,d/2)$. Thirteen 
photon energies varying from 100~MeV to 20~GeV are generated, leading to a total of 169 
configurations. In each case, the response on the calorimeter is studied for $10^{5}$ generated initial photons: the event-by-event 
energy deposit in each individual block is recorded and used to reconstruct the shower central position ($x_{c}$,$y_{c}$) as 
described in the following sections. Focus is put here only on the $x_{c}$ coordinate since the $y_{c}$ coordinate can be deduced 
following the same methods.

Two approaches can be used for reconstructing the shower central position $x_{c}$. The first method is based on a 
numerical formula depending on the $x_{i}$-coordinate of the $i^{th}$ block center and the energy deposit $E_{i}$ in this block. 
The second method developed in the present work relies on fitting the calorimeter block response with a known 
profile function where $x_{c}$ becomes the free parameter of the fit.


\section{Formula based method}
\label{sec3}

Various formulas have been proposed to determine the shower central position~\cite{{ref7},{ref8},{ref9},{ref10},{ref11}}. 
The center of gravity method is one of the most common formula
\begin{equation}
x_{c} = \frac{\sum_{i}w_{i}x_{i}}{\sum_{i}w_{i}}~~, 
\label{eq1}
\end{equation}
where the sum runs over the number of blocks in the shower cluster and $w_{i}$ is a weight factor depending of the energy $E_{i}$. 
In the simplest case where $w_{i}=E_{i}$, it has been shown that the obtained $x_{c}$ depends on $x$ in a non-linear way. A term, 
depending of the size $d$ of the blocks and the exponential radial falloff of the shower, has to be added to Eq.~\ref{eq1} to 
correct from the correlation between $x_{c}-x$ and $x$~\cite{ref7}, the so-called {\it S-curve}. Many other formulas and algorithms 
have been discussed and compared in reference~\cite{ref9}. For example, the un-shifted estimate of $x_{c}$ based on here-after 
expressions (Eq.~\ref{eq2}-\ref{eq3}) reduces significantly the non-linear correlation between $x$ and $x_{c}$ and gives a satisfying reconstruction of the shower position
\begin{equation}
x_{c} = x_{m}+\frac{\sigma_{1}}{2}\ln{\left(\frac{E_{m+1}}{E_{m-1}}\right)}~~, 
\label{eq2}
\end{equation}
\begin{equation}
x_{c} = x_{m}\pm d \mp \sigma_{2}\ln{\left(\frac{1}{2}\left[1+\frac{E_{m}}{E_{m\pm1}}\right]\right)}~~. 
\label{eq3}
\end{equation}
In the previous relations $x_{m}$ represents the coordinate of the block with a maximal energy deposition and $\sigma_{1}$ and 
$\sigma_{2}$ are related to the transverse exponential behavior of the shower~\cite{ref9}. This un-shifted estimate of $x_{c}$ uses 
only the information of two particular blocks: the ones with the largest energy deposition (Eq.~\ref{eq3}) and the ones adjacent to 
the block with maximal energy (Eq.~\ref{eq2}). In comparison to the simplest energy weighting case, the logarithmic weighting of 
energies in these  expressions procures a stronger influence of low energy deposit blocks in the calculation of $x_{c}$. Lately, a 
simple method based on Eq.~\ref{eq1} has been proposed and is still largely employed nowadays in many experimental analysis. It has 
been shown that it gives results similar or superior in quality to all those discussed previously and does not need position correction~\cite{ref11}. In this method the energy weight $w_{i}$ is given by
\begin{equation}
w_{i}=\max \left\{ 0; W_{0}+\ln\left( \frac{E_{i}}{E} \right) \right\} 
\label{eq4}
\end{equation}
where $W_0$ is a free dimensionless parameter. Consequently, only the blocks having an energy deposition higher than $E e^{-W_0}$ are taken into 
account in the calculation of $x_{c}$. In addition $W_0$ allows to set the relative weight of the blocks, with a small energy 
deposition, used in the sum of Eq.~\ref{eq1}. Indeed, high $W_0$ values attribute almost an equal weight to the blocks entering the 
sum, while small $W_0$ values favor the highest energy blocks. It exists therefore an optimal $W_0 \approx 4$  giving the best 
position reconstruction~\cite{ref11}. 

\begin{figure}[h]
\begin{center}
\epsfig{file=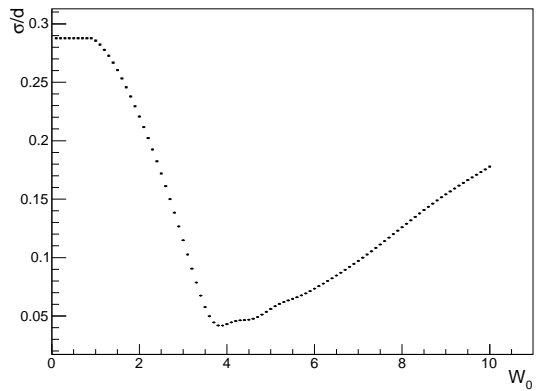, width=8cm}
\vspace*{-10pt}
\caption{Relative position resolution $\sigma/d$, for $d=r_M$ and $E=10$ GeV, as a function of $W_0$ (statistical  uncertainties are smaller than the point size).}
\label{fig2}
\end{center}
\end{figure}
\begin{figure}[t]
\begin{center}
\epsfig{file=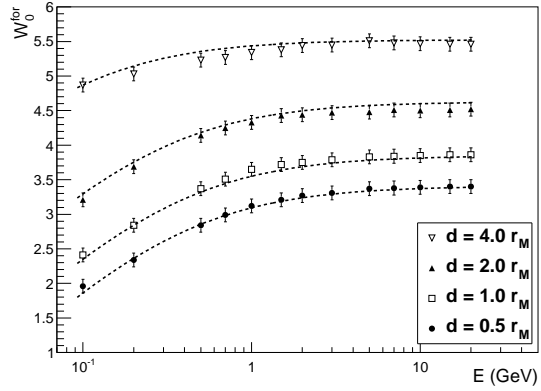, width=8cm}
\vspace*{-10pt}
\caption{Energy dependence of $W_0^{for}$. The block size $d$ relative to the Moli\`ere radius is reported in the legend 
for each configuration. Dashed lines represent the parametrization of Eq.~\ref{eq5}. The 0.1 uncertainty on $W_0^{for}$ corresponds to the bin size of Fig.~\ref{fig2}.}
\label{fig3}
\end{center}
\end{figure}

Actually the optimal $W_0$ value, called $W_0^{for}$ hereafter to indicate the formula method origin of this parameter, depends on 
the size $d$ of the blocks as well as the energy $E$ of the incident particle. Fig.~\ref{fig2} shows a typical behavior of the position resolution $\sigma$, defined by the root mean square (RMS) of the $(x_c-x)$ distribution, as a function of $W_0$. The minimal value of 
$\sigma$ defines $W_0^{for}$ whereas the worst resolution obtained for $W_0=0$ or $W_0\rightarrow+\infty$ equals $d/\sqrt{12}$ and corresponds to the RMS of a uniform distribution of width $d$. The optimal $W_0^{for}$ and corresponding position resolution $\sigma^{for}$ are determined for each geometry and energy configuration  previously described (Sec.~\ref{sec2}). Fig.~\ref{fig3} shows the energy dependence of $W_0^{for}$ for different block sizes. At a given size, the energy deposit in the blocks surrounding the central block becomes small and very sensitive to the  sampling fluctuations of the shower when the energy $E$ decreases. Including these blocks in the calculation of $x_c$ could then degrades the resolution. These blocks are removed by small $W_0$ values as shown on Fig.~\ref{fig3}. At given energy $E$, $W_0^{for}$ increases as the block size $d$ becomes larger. Indeed, for large blocks the energy of the shower is essentially deposited in the central block. Having a high threshold with small $W_0$ excludes the remaining blocks and the $x_c$ position becomes the central block coordinate with a resolution relative to the block size. An empirical parametrization of $W_0^{for}$ as function of $d$ and $E$ is proposed here following the expression 
\begin{equation}
W_0^{for}=\ln \left(\frac{100~E(\mathrm{GeV})}{2.01~e^{-\frac{d}{r_M}} + \left[4.95~e^{-\frac{d}{r_M}}+0.307\right]E(\mathrm{GeV})}\right) 
\label{eq5}
\end{equation}
where the three numerical constants are determined from a global fit of the 169 configurations. Fig.~\ref{fig3} shows the result of this parametrization for some particular values of $d$.

\begin{figure}[t]
\begin{center}
\epsfig{file=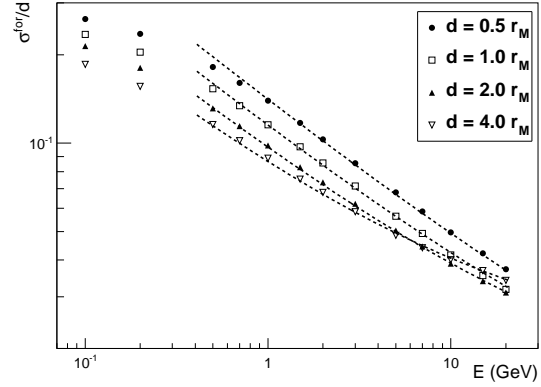, width=8.cm}
\vspace*{-10pt}
\caption{Energy dependence of the relative resolution. Dashed lines represent Eq.~\ref{eq8} parametrization.}
\label{fig4}
\end{center}
\end{figure}

\begin{figure}[h]
\begin{center}
\epsfig{file=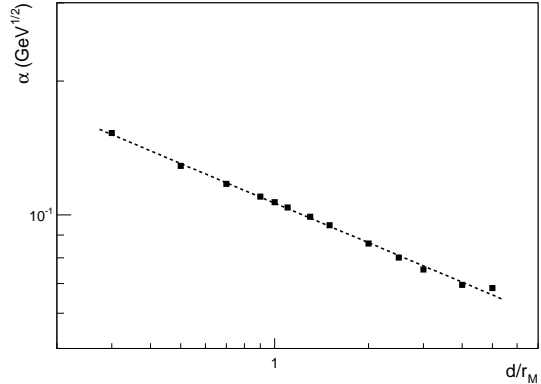, width=8.cm}
\vspace*{-10pt}
\caption{The fitted coefficient $\alpha$ as a function of $d$ (in $r_M$ units).}
\label{fig5}
\end{center}
\end{figure}

The obtained resolution $\sigma^{for}$ is represented on Fig.~\ref{fig4} as function of $E$ for different block sizes. 
$\sigma^{for}$ is expected to scale with $1/\sqrt{E}$ at high energies~\cite{ref12}: the relative energy 
resolution $\sigma_i/E_i$ for each block $i$ is proportional to $1/\sqrt{E_i}$ and then to $1/\sqrt{E}$, assuming 
the same average shower profile; all the terms in Eq.~\ref{eq1} having then a relative precision 
proportional to $1/\sqrt{E}$, $\sigma^{for}$ is also scaling as $1/\sqrt{E}$. At low energies ($E\ll$ 1~GeV), 
this proportionality becomes not valid since the worst resolution one could obtain cannot exceed $d/\sqrt{12}$. 
The $E$-dependent $\sigma^{for}$ graphs for each block size are fitted at high energies with the expression 
\begin{equation}
\frac{\sigma^{for}}{d}=\frac{\alpha}{\sqrt{E}}+\beta~~, 
\label{eq6}
\end{equation}
where $\alpha$ and $\beta$ are the free parameters of the fit. The resulting coefficient $\alpha$ is represented on 
Fig.~\ref{fig5} as function of the block size. The $d$-dependence of $\alpha$ can be parametrized as 
\begin{equation}
\alpha=a^{for}\left(\frac{d}{r_M}\right)^{b^{for}}
\label{eq7}
\end{equation}
where $a^{for}$ and $b^{for}$ are two constants relative to the formula method. The global fit of the $\sigma^{for}$ 
resolution for each configurations leads to the semi-empirical parametrization of the relative resolution 
\begin{equation}
\frac{\sigma^{for}}{d}=\frac{a^{for}\left(\frac{d}{r_M}\right)^{b^{for}}}{\sqrt{E(\mathrm{GeV})}}+c^{for}\left(\frac{d}{r_M}\right)+d^{for}
\label{eq8}
\end{equation}
where $a^{for}=0.110$, $b^{for}=-0.334$, $c^{for}=3.55 \times 10^{-3}$ and $d^{for}=4.02 \times10^{-3}$. The results 
of this expression are shown in Fig.~\ref{fig4} for some particular block sizes. It is obvious from Eq.~\ref{eq8} that 
the resolution $\sigma^{for}$ becomes better at high energies and for small block sizes. However, the $d$-dependence of 
the  relative resolution $\sigma^{for}/{d}$ at fixed energy is more involved. It is worth noting that Eq.~\ref{eq8} suggests 
the existence of an optimum block size corresponding to an optimum relative resolution at a given $E$, allowing optimizing 
the calorimeter block size for a given energy measurement range.
 

\section{Fit based method}
\label{sec4}
 
\begin{figure}[h]
\begin{center}
\epsfig{file=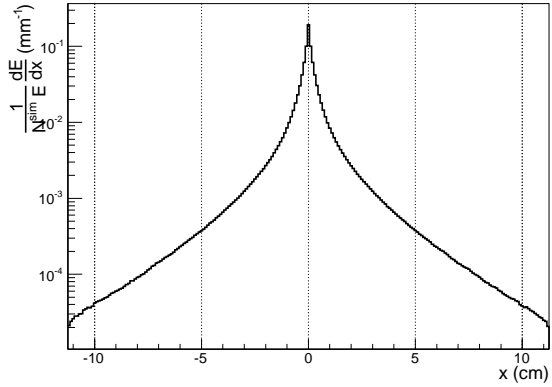, width=8.cm}
\vspace*{-10pt}
\caption{Average shower profile function for $E$=1~GeV photons.}
\label{fig6}
\end{center}
\end{figure}
A second approach for reconstructing the shower central position relies on the knowledge of the lateral profile of the 
electromagnetic shower. Fig.~\ref{fig6} shows the profile function $F(x)$ obtained from the simulated response of an infinite 
lead fluoride calorimeter to $N_{sim}$=$10^5$ photons of 1~GeV. This response is represented as function of the distance 
from the shower center along $x$-axis. The energy deposit is here determined per generated photon, as the transverse dimension 
integral normalized by the initial photon energy 
\begin{equation}
F(x) = \frac{1}{N_{sim}} \, \frac{1}{E} \, \int_{-\infty}^{+\infty} \frac{d^{2}E(x,y)}{dxdy} \, dy = \frac{1}{N_{sim}}\, \frac{1}{E} \, \frac{dE(x)}{dx}
\end{equation}
where $dE(x)$ is the energy released by the shower in a vertical column calorimeter centered at $x$, having a width $dx$=1~mm 
and an infinitely large height. This profile function turns out to be the same for photons and electrons and no energy dependence 
is expected according to the definition of the Moli\`ere radius.

In Ref.~\cite{ref9}, the shower profile was approximated by a single exponent term and a least-squares fit algorithm was  developed 
for extracting the shower position taking into account all of the responding blocks of the calorimeter. It was  shown that the results 
are similar to those obtained with Eq.~\ref{eq2} and Eq.~\ref{eq3}~\cite{ref9}, and do not present any significant advantage relatively 
to the fast and simple method based on Eq.~\ref{eq1}-\ref{eq4}~\cite{ref11}. Nevertheless, it is clear from Fig.~\ref{fig6}, that the 
lateral profile of the shower is a more intricate function different from a simple exponent or a combination of two exponent contributions. 
A more realistic form $F(x)$ of the profile function is proposed in the present work, deduced from the bin-to-bin linear interpolation of 
the simulated profile function (Fig.~\ref{fig6}). 

In the calorimeter described in section~\ref{sec2}, the expected energy deposit for a given simulated event in a column $i$ identified by its central coordinate $x_i$ writes 
\begin{equation}
E_i^{exp}= E \, \int_{x_i-d/2}^{x_i+d/2} \, F(x-x_c) \, dx
\label{eq9}
\end{equation}
where $x_c$ is the unknown shower central position, and $E$ is the incident particle energy corresponding to the total energy deposit in the calorimeter if we neglect the energy resolution effect. 
The clear energy deposit in that column for the same event expresses
\begin{equation}
E_i=\sum_{j}E_{ij} \,\, \Theta(E_{ij}-E~e^{-W_0})
\label{eq10}
\end{equation}
where $W_0$ is a dimensionless parameter related to the energy threshold applied on the blocks, and $\Theta(x)$=1 when $x>0$ and $0$ 
otherwise is the Heaviside function. In the previous equation, $E_{ij}$ represents the energy deposit in the block belonging to column $i$ and row $j$. 
Fitting for each event, with the MINUIT package~\cite{ref13}, the $E_i$ distribution (Eq.~\ref{eq10}) with the expected distribution $E_i^{exp}$ (Eq.~\ref{eq9}) allows 
to extract the single free parameter of the fit $x_c$. Following Eq.~\ref{eq10}, this algorithm rejects the blocks with energy deposit 
smaller than the threshold $E~e^{-W_0}$. 

\begin{figure}
\begin{center}
\epsfig{file=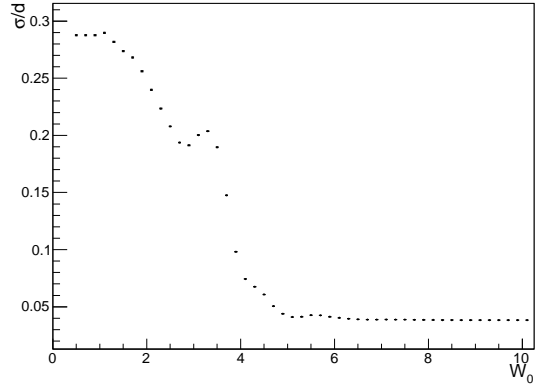, width=8cm}
\vspace*{-10pt}
\caption{$W_0$-dependence of the relative position resolution $\sigma/d$, for $d$=$r_M$ and $E$=10~GeV.}
\label{fig7}
\end{center}
\end{figure}
Fig.~\ref{fig7} shows the influence of $W_0$ on the obtained resolution $\sigma$ given by the RMS of the $(x_c-x)$ distribution for a 
particular value of $E$ and $d$. As expected, larger the block number contributing to Eq.~\ref{eq10}, better the relative resolution. 
However, the resolution remains constant after a certain $W_0$ since very small energy  deposit blocks do not contribute significantly 
to $x_c$ calculation and consequently not change the fit result. The optimum $W_0$, denoted $W_0^{fit}$ here-after, is defined as the 
value above which the resolution does not improve by more than 1\%.
\begin{figure}
\begin{center}
\epsfig{file=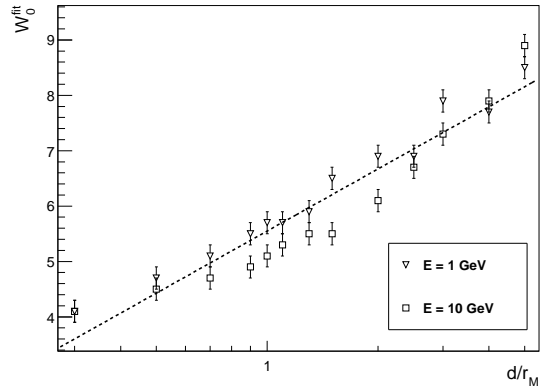, width=8cm}
\vspace*{-10pt}
\caption{Block size dependence of $W_0^{fit}$ for two different photon energies; the dashed line corresponds to  Eq.~\ref{eq11} parametrization.}
\label{fig8}
\end{center}
\end{figure}
No real energy dependence of $W_0^{fit}$ is observed but logarithmic scaling with the block size is demonstrated on Fig.~\ref{fig8}, 
following the parametrization
\begin{equation}
W_0^{fit}=1.61~\ln\left(\frac{d}{r_M}\right)+5.55~~.
\label{eq11}
\end{equation}
For large block size, the shower energy is mainly deposited in one single block so the energy threshold must be reduced to 
include the surrounding blocks in order to improve the spatial resolution. 

\begin{figure}[h]
\begin{center}
\epsfig{file=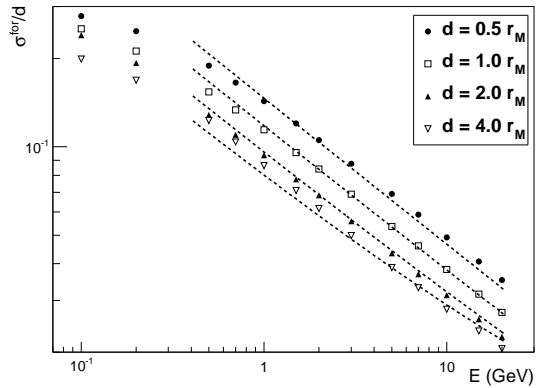, width=8cm}
\vspace*{-10pt}
\caption{The relative resolution $\sigma^{fit}/d$ as a function of the shower energy $E$. The block size $d$ is indicated 
in the legend for each configuration. Dashed lines represent the parametrization of Eq.\ref{eq12}.}
\label{fig9}
\end{center}
\end{figure} 
The relative resolution of the fit method here developed is represented in Fig.~\ref{fig9} as function of the shower energy 
$E$ for different block sizes. As for the formula method, the resolution can be parametrized following the expression
\begin{equation}
\frac{\sigma^{fit}}{d}=\frac{a^{fit}\left(\frac{d}{r_M}\right)^{b^{fit}}}{\sqrt{E(\mathrm{GeV})}}+c^{fit}\left(\frac{d}{r_M}\right)+
d^{fit}~~, 
\label{eq12}
\end{equation}
where $a^{fit}=0.121$, $b^{fit}=-0.349$, $c^{fit}=1.78 \times 10^{-3}$ and $d^{fit}=1.89 \times 10^{-3}$. These four constants 
are deduced from a global fit of the studied configurations. 

The fit algorithm described in this section is obviously more computing time consuming at the data analysis level than the 
formula based method. Nevertheless, modern computers can easily and efficiently handle this problem: for instance, more than 
$10^4$ events per second can be analyzed with a modest 2~GHz processor. 


\section{Discussion}
\label{sec5}

\subsection{Comparison of the formula and fit methods}
\label{sec5.1}

\begin{figure}[h]
\begin{center}
\epsfig{file=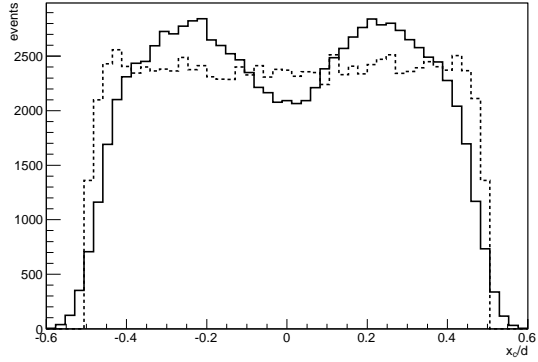, width=8cm}
\vspace*{-10pt}
\caption{Reconstructed shower position $x_c$ (in block size units) using the formula method (solid line) and the fit method (dashed 
line) for $d=r_M$ and $E=5$~GeV.}
\label{fig10}
\end{center}
\end{figure}
The formula and fit methods are compared on Fig.~\ref{fig10} from a typical distribution of the shower central position obtained for a 
given configuration. In opposition to the fit determination, the formula determination of $x_c$ does not exhibit the constant behavior 
expected from the uniform generation of the shower $x$ origin. This appears as a consequence of the number of blocks taken into account 
in Eq.~\ref{eq1}. It was found in this study that the average number of blocks entering Eq.~\ref{eq1} is approximately 4 but could 
slightly differ depending on the configuration type (for instance, 4.2 in the configuration shown on Fig.~\ref{fig10}). Therefore on 
average, two different block coordinates only are contributing to the determination of $x_c$ in Eq.~\ref{eq1}, which slightly favors the region 
between the center and the boundaries of the central block. This restriction does not show-up in the fit method where $W_0^{fit}$ is 
always larger than $W_0^{for}$ and allows a larger number of blocks to contribute to the determination of the shower central position.

\begin{figure}[h]
\begin{center}
\epsfig{file=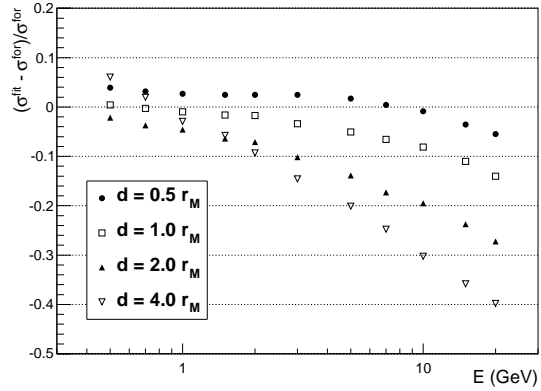, width=8cm}
\vspace*{-10pt}
\caption{Energy dependence of the relative difference between the formula and fit methods. The block size $d$ is indicated in the legend 
for each configuration.}
\label{fig11}
\end{center}
\end{figure}
Fig.~\ref{fig11} shows the relative difference between the spatial resolution obtained with the two previously described methods. For small block sizes, the formula based method tends to be slightly preciser than the fit based method. However, when the block size comes closer or higher than the Moli\`ere radius, the fit based method can provide as high as 40\% better resolution at high energies. Within the present work, particles with normal incidence only are considered whereas experimentally one can have different incident angles. It has been shown that the position resolution obtained with the formula based method is rather insensitive to this angle for moderate values~\cite{ref11}. For large  incidence angles, a simulation optimizable geometrical correction depending on the shower depth must be taken into account since the impact position of the particle is shifted relatively to the shower central position~\cite{ref15}. Similarly, different lateral shower profile functions depending  on the incidence angle can be simulated and exported within the fit method to obtain the shower position and the corrected impact position.

\subsection{Effect of the energy resolution}
\label{sec5.2}

The resolution of the energy measurement per block originates solely from the shower sampling fluctuations in the current simulation approach. Experimentally, lead-fluoride calorimeters are based on the detection of \v{C}erenkov light and the fluctuations in the number of collected  photo-electrons dominates the energy resolution~\cite{ref12}. The generation and tracking of \v{C}erenkov photons is not performed here because of unrealistic computing times and strong sensitivity to exact optical properties of crystal and wrapping surfaces which are known to differ  from an experimental device to another. This effect can globally be symbolized by an additional smearing of the energy deposition $E_i$ in each block $i$. This is done by adding to $E_i$ a random number following a Gaussian distribution centered at zero and having the following width 
\begin{equation}
\sigma^{s} = s \, \sqrt{E_i} \,\, ,
\label{eq13}
\end{equation}
where $s$ is a constant related to the global relative energy resolution of the calorimeter. Several $s$-values in the range 1\%-20\% are used in the following to study the effect of the energy resolution on the position resolution. The degradation of the position resolution $\sigma$ is defined as  
\begin{equation}
\sigma^{deg} = \sqrt{\sigma^2(s)-\sigma^2(s=0)}
\label{eq14}
\end{equation}
where $\sigma(s)$ is the position resolution at a given $s$-value, and $\sigma(s$=$0)$ is the position resolution in absence of energy 
smearing, both determined for any of the two methods. This study shows that $\sigma^{deg}$ is proportional to $s$. 

\begin{figure}
\begin{center}
\epsfig{file=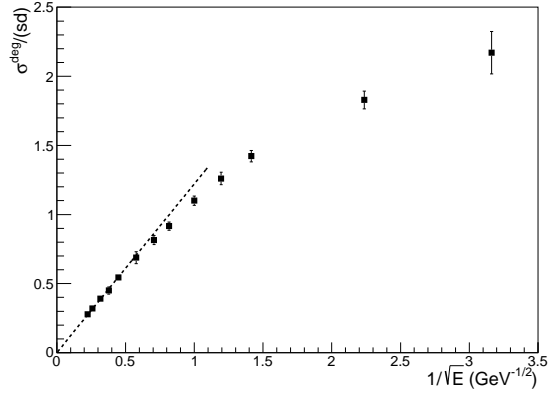, width=8cm}
\vspace*{-10pt}
\caption{$\sigma^{deg}/s/d$ as a function of $1/\sqrt{E}$ for the $d=r_M$ configuration. The same behavior is observed for other 
configurations which are omitted sake of clarity.}
\label{fig12}
\end{center}
\end{figure}

\begin{figure}
\begin{center}
\epsfig{file=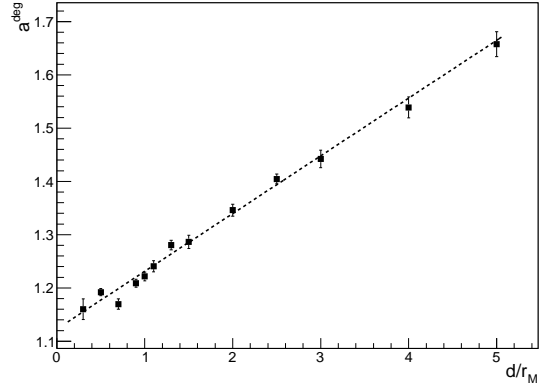, width=8cm}
\vspace*{-10pt}
\caption{The fitted parameter $a^{deg}$ (Eq.~\ref{eq15}) as a function of $d/r_M$. The dashed line corresponds to the relation 
$a^{deg}$=$ 0.112 \left(\frac{d}{r_M}\right)+1.12$ and represents the first order polynomial fit of the distribution.}
\label{fig13}
\end{center}
\end{figure}

Fig.\ref{fig12} shows the typical energy dependence of the relative average resolution $\sigma^{deg}/s/d$ for a particular block size. 
As expected from Eq.~\ref{eq13}, the degradation of the position resolution can be parametrized at high energies 
\begin{equation}
\frac{1}{s} \, \frac{\sigma^{deg}}{d} = \frac{a^{deg}}{\sqrt{E}} 
\label{eq15}
\end{equation}
where $a^{deg}$ is a parameter depending linearly on the block size $d$ as shown on Fig.~\ref{fig13}. Finally the obtained position 
resolution $\sigma$ taking into account energy resolution effects can be expressed
\begin{equation}
\frac{\sigma}{d} = \frac{\sigma^{for,fit}}{d} \oplus ~s~\frac{0.112\left(\frac{d}{r_M}\right)+1.12}{\sqrt{E(\mathrm{GeV})}} 
\label{eq16}
\end{equation}
where $\sigma^{for,fit}$ is either given by Eq.~\ref{eq8} or Eq.~\ref{eq11} according to the chosen method of the position determination.

Experimentally, other effects such as the physics and electronics background, energy calibration, radiation damage... can contribute to 
the energy resolution and then degrades the position resolution. However, in a well designed experiment these effects are not dominating the  position resolution. Finally, the systematic uncertainty coming from the experimental knowledge of the coordinates $x_i$ of the block centers has to be added quadratically to Eq.~\ref{eq16} to obtain the final position resolution but this contribution can generally be neglected, as connected to the accuracy of the mechanical design and mounting. In Ref.~\cite{ref3}, the obtained experimental resolution on the 
reconstructed position using the center of gravity method is 2~mm, in good agreement with the 1.9~mm predicted by Eq.~\ref{eq16}.


\section{Conclusion}
\label{sec6}

The present work discussed the determination of the shower central position in a laterally segmented electromagnetic calorimeter 
following the widely used center of gravity method and a new fit method here developed. A semi-empirical parametrization of the 
relative position resolution as a function of the incident particle energy and the calorimeter block size is proposed and 
optimized for each case. The fit method is shown to significantly improve the position resolution as compared to the formula 
method, particularly at high energies and for large block sizes. Finally, energy resolution effects on the position resolution 
are also discussed and quantified. 

Despite the present application to lead-fluoride calorimeters, the results presented here can be exported to several electromagnetic 
calorimeter materials through their parametrization in terms of the Moli\`ere radius. In addition, as far as simulations are 
concerned, the profile function at the heart of the fit method has energy and material universality features. This global study 
not only provides useful parametrization for the design optimization of electromagnetic calorimeters in future experiments looking  
for a good spatial resolution, but procures also a new algorithm to improve position resolution at the data analysis level.   



\end{document}